# A comparative study of machine learning techniques used in non-clinical systems for continuous healthcare of independent livings

Zahid Iqbal[1], Rafia Ilyas[2], Waseem Shahzad[3], Irum Inayat[4]
[1,2]Department of Computer Science, University of Gujrat, Gujrat, Pakistan
[3,4]Department of Computer Science, FAST University, Islamabad, Pakistan
Email: [1]zahid.iqbal@uog.edu.pk, [2]rafy.choudary@gmail.com, [3]Waseem.shahzad@nu.edu.pk ,
[4]irum.inayat@nu.edu.pk

*Abstract*—New technologies are adapted to made progress in healthcare especially for independent livings. Medication at distance is leading to integrate technologies with medical. Machine learning methods in collaboration with wearable sensor network technology are used to find hidden patterns in data, detect patient movements, observe habits of patient, analyze clinical data of patient, find intention of patients and make decision on the bases of gathered data. This research performs comparative study on non-clinical systems in healthcare for independent livings. In this study, these systems are sub-divided w.r.t their working into two types: single purpose systems and multi-purpose systems. Systems that are built for single specific purpose (e.g. detect fall, detect emergent state of chronic disease patient) and cannot support healthcare generically are known as single purpose systems, where multi-purpose systems are built to serve for multiple problems (e.g. heart attack, fall detection etc.) by using single system. This study analyzes usages of machine learning techniques in healthcare systems for independent livings. Answer Set Programming (ASP), Artificial Neural Networks, Classification, Sampling and Rule Based Reasoning etc. are some state of art techniques used to determine emergent situations and observe changes in patient data. Among all methods, ASP logic is used most widely, it is due to its feature to deal with incomplete data. It is also observed that system using ANN shows better accuracy than other systems. It is observed that most of the systems created are for single purpose. In this work, 10 single purpose systems and 5 multi-purpose systems are studied. There is need to create more generic systems that can be used for patients with multiple diseases. Also most of the systems created are prototypical. There is need to create systems that can serve healthcare services in real world. Some systems are hard to be used in real life due to large hardware requirements. Although, good systems are created but still there is need to build more efficient, affordable, adoptive and generic systems.

## I. Introduction

About 20% of world population will have age 60 or above. Population of older is increasing than children [1]. Aging is bringing health related problems e.g. increase in diseases, increase in healthcare cost and shortage of care givers etc. To overcome these problems non-clinical healthcare systems get attention. Non clinical systems are built for independent livings. Fast advances in sensors and networking technologies and their integration with intelligent systems make it possible to get personal healthcare services at distant [1]. Remarkable advantages are taken from machine learning methods [3-7]. Non-clinical systems require continuous and fast data collection and processing to get useful information [2]. In healthcare, data may contains patient profile (history), values of vital signs and medical data. For this sake, machine learning techniques are playing most vital role in healthcare research. It is hard to analyze data and draw important conclusions from it. Different machine learning techniques [2] are being used to analyze complex patient's data to extract useful information. These techniques analyze patient state and detect emergent situations [8-10] and response accordingly. The main aim of using intelligent systems is not only to detect problem but also provide solution to that problem. This review will analyze selected systems proposed in past few years for independent livings. These systems detect emergent situations [8], do risk analysis, provide personal care [11], work as a coach and inform how to response under particular scenario. This research analyzes advantages and limitations of these systems. This research analyzes state of art techniques used in these systems.

## II. Systems and Techniques

Several systems are proposed in last few years for continuous healthcare by using machine learning technique. This research study systems that have following features: a device for collecting data; an "intelligent" part to process on patient's data; Ability to detect certain situations; response after detection and system type can be prototypical, implemented, proposed model etc. To increase level of understanding we categorized these systems into two major categories: Single Purpose and Multi-Purpose systems.

**A.** *Single Purpose Systems:* Systems that are built for specific disease are called single purpose. Special systems are made for fall detection [12-14]. In these systems, ambient intelligence

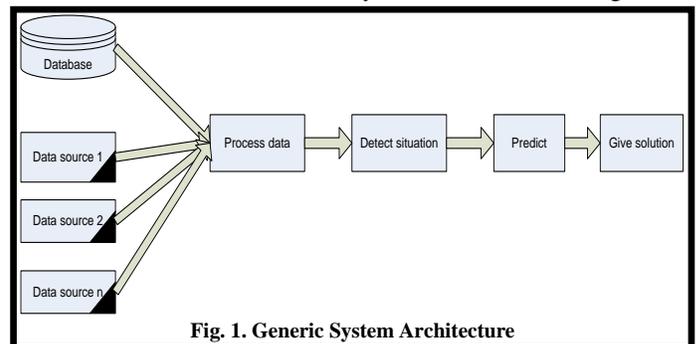

Fig. 1. Generic System Architecture

with single and multiple agents are used. Architecture of a generic system is given in Fig 1.

    i. Ambient intelligence is an electronic environment built up of smaller devices connected with each other and is responsi*ve to people. Single* ambient agent is used as a personal coach to guide patient about its exercise actions [15]. This personal coach observes movements of a person and takes record of its normal movements and time for that movement.

| | | | |
|---|---|---|---|
| **A P P L I C A T I O N S** | Single purpose | Ambient Intelligence | **1.** M healthcare |
| | | | **2.** Personal coach |
| | | **3.** I.T CARE | |
| | | Emergency Detection | **4.** Ontology based model |
| | | | **5.** Chronic disease emergency detection |
| | | | **6.** emergency detection using smart phone |
| | | | **7.** Chronic disease emergency detection |
| | | Fall Detection | **8.** Vision based |
| | | | **9.** Wearable based |
| | | | **10.** Ambient based |
| | Multi-purpose | Context Awareness | **11.** Context care |
| | | | **12.** Monitoring contextual situation |
| | | | **13.** Home healthcare |
| | | | **14.** Reasoning to detect emergency |
| | | | **15.** Cyber-care |

**Fig. 2. Application systems**

Coach learning and set of normal movement vary from person to person. Coach then compare present movement of a person with real movement to identify gaps. This learning coach can be used for exercise and recovery after some incident [15]. Minor changing in physical movement of a person will also be detected as false movement. On the other hand, physical movements vary from person to person.

And multi-agents are used for online interaction of patient and doctor. It also takes doctors feedback on emergent situations Six major agents are used to build system [16]. Every agent performs its specific job and is connected to agent next to it. First agent gathers data, second agent check and pass it, third agent filters received data and send it to agent connected to doctor. Then doctor response back to situation. Data is stored in online databases. Major decision making task is performed by sixth agent. For analyzing data association rule mining [17] is used. Historical data of patient is also used. It make to classes of data "normal" and "emergency". And check every situation under these two classes [16]. It also updates patient record. System is inexpensive and flexible in communication. But large memory and energy is requirement of system.

 ii. I.T- based services

A system that provides tailoring facility is named U-care [11]. Care-givers IT-based tailoring means that care-giver can set their service plans (e.g. alarms for medicine, service alerts, and calendars etc.) and also can make decision on the basis of professional knowledge on distance. Alerts are sent to care-receiver on its mobile phone [11]. System is inexpensive and it improves quality of life. But as it is created for elderly, it is difficult for elderly to use and understand system functions. Usually, elderly people do not like to use such systems.

 iii. Emergency Detection in Chronic Disease Patients
Several systems are built to detect emergent situations for independent livings. These systems continuously monitor independents. Systems are specially built to detect emergency for chronic disease patient. Some Emergency detection systems are: (ontology based, real time health care, CRNT+, HUMECS and fall detection).

Formal representation of knowledge in form of hierarchy of concepts within an area using common properties, types and interrelation of concepts is called ontology [18]. An ontology-based context-aware model for patients with chronic conditions is proposed in [18]. This system is proposed to provide long term health services. System collects data that contains patient's biomedical data, location, environmental data and social data. Social data is taken because patients with chronic conditions have greater effect of society. Reasoning is performed over knowledge (gain from ontology based formulism) and knowledgebase using rule-based and ontology-based engines to detect risk related to patient's conditions [18]. Accuracy of a system can be considered 64%. But wrong data entry can make solution worse. Huge data is required for [18].

A real time healthcare using classification and association rule mining [17] model is proposed for chronic disease patients [10]. System takes clinical, historical and continuous data of patient. To predict patient condition, "data filtering and feature extraction is performed to remove unwanted data factors. Major flaw of rule mining is that increasing data will increase rules that will cause increase in processing time.

A toolbox named CRNT+ is developed by [9] for smart phones. This system analyze whether patient is in need of external help or not. Sensors are fixed inside phone to collect patient's data, after collecting data, sampling and classification [17]] is performed over that data. If external help is needed system send alarms of help to caregivers. Process is performed in real time. Data about patient is stored online and fetch further when needed. This system is extended form of CRNT [19]. CRNT+ takes low energy and it is extendable. Major drawback of CRNT + is that works only in day times.

Hoseo University mobile e-healthcare system (HUMECS) using Neural Network is proposed to deal with patients that needs first aid. HUMECS provides mobile based E-healthcare that detects location of patient in emergency and send message to guardian and hospital to provide emergency measures. For analyzing patient's data Personal mobile host (PMH) is used. PMH has different sensor modules (acceleration sensor, vibration sensor, GPS). Data related to normal walking, fainting and seizures etc. is collected from sensors and inputted in back propagation network. Network diagnoses whether patient is in emergency or not? If emergency is detected than message is sent to control center that includes patient's information and location. [8]. But as data will be gathered from far points, it may holds noise which can cause wrong detection.

 iv. Fall detection

Fall detection systems are very important especially in elderly care. It is observed from a survey that death rate due to "falls" is increasing. Three third of people have age above 65 died due to "falls" [20]. There are three main categories of fall detection systems. Following are categories: [14] ambient based, vision based and wearable based.

*Systems* that depend on pressure sensors are called ambient based systems. A non-wearable ambient based system to detect fall is proposed in [12]. This system use Doppler-Based sensors to collect features. Two classification [17] methods (Support vector machine, K-nearest neighbor) are used to classify extracted features. System gives alarms to detect fall [12]. System is checked in close room. Where there was no interference. Structure of system shows that it may not work well in open area. As outer interference will be there.

Systems that are related to shapes and postures of body are called vision based systems. A vision based system that collects full human posture in 3D form is proposed in [13] to detect fall. Preprocessing is performed over data by using Principal component Analysis (PCA). A database is used to store all possible moves of fall from several different people. And compare user moves with stored moves. System is inexpensive and result of system is 92%-100% [013]. But with increasing data system performance will be decrease.

Methods that used wearable sensors to detect fall are called wearable based methods [14]. A wearable sensors based system that uses Signal Measurement Vector (SMV) to detect falls effectively is proposed in [14]. This system gives accuracy of 96%. It uses batteries that consume low power. But at the same time it is expensive to implement such embedded systems. It is difficult for a patient to handle wires and wear sensors all the time. Position of sensor will affect output of system.

*B. Multi-purpose systems:* Systems that are built to serve multiple problems (heart attack, fall detection etc.) fall under multi-purpose systems. These systems can support in several situations (heart attack, fall detection etc.).

   a. Emergency Detection through Context Awareness

Context Awareness is property of a system to understand current location and condition of user and response according to the contextualized condition [21]. In healthcare, context awareness in used to identify the condition of patient and his interaction with application. Different applications of healthcare with context aware ability are: Cyber care, SINDI, home healthcare, context care,

A system named Cyber Care [22] collected patient clinical data through wearable acquisition devices and movements of patient through sensors in proposed in [22]. ASP logic program is implemented to detect emergency from data. This system deals with continuous real time data. Both static and dynamic data is collected. Static data holds clinical data. Dynamic data holds patients biomedical parameters, list of moves, patients. Data is converted into logic predicates. This program converts solution into actions [22].

A logical approach for home health care is proposed in [23]. Reasoning component based on ASP logic program is used for detecting patient situation. ASP logic performs three major reasoning tasks: contextualizing patient's physical, mental and social state continuously, predicting possible risky situation and finding reasons for risky situations. On every inference, indicators are compared with previous inference. Different evaluations are performed. This system results 88% accuracy. But settlement of sensors can effects data input also moves of hands and arms cannot be detected [23].

A contextual model that based on ASP logic is introduced in [24] for situation judgment. This model is highly declarative for feed-back policy. Logic base of this system is used to: Understand risky situations in context awareness way and give feed-back according to situation.

Contexta-care [25] system that collects continuous physical data of patient and predict risky situations using rule based is proposed in [25]. Three type of rules are generated: Rule of physical movements, Rules for environment and Rules for both feature [25]. But large memory is required for this system. Increase in data will decrease efficiency of system. System may require more time for rules making.

A system with intelligent reasoning component is proposed in [26]. Intelligent logical reasoning] component collects continuous contextual, mental and social state of person, analyze that data and predict risky situation. It also identifies reason of emergency [26]. Master processor is used for fast processing and to fulfill memory requirements. But it makes system costly and difficult to implement.

### III. TECHNIQUES

From above section it is analyzed that several machine learning techniques [2] are used in healthcare to provide services to individual. This section will give introduction of most randomly used techniques. Machine learning is sub-discipline of Artificial Intelligence. Machine learning techniques have ability to learn from heuristic data (training data) and predict accordingly [27]. Following machine learning techniques are used in above discussed systems.

i. Artificial Neural Networks (ANN)*:* Model of neural networks hold set of interlinked nodes. A unit that take some input and produce useful output is called node. Values of nodes are determined by their links. Back propagation Neural Network is basic type of ANN [27]. Values are propagated back till it meets specific threshold [8]. In healthcare back propagation neural network is used in [8].

ii. Support vector Machine*:* Support Vector Machine (SVM) is supervised machine learning technique. SVM is used to identify patterns in data and then classify it accordingly.

iii. Signal measurement vector: measurement vector (SMV) is a machine learning technique used in signal processing.

iv. Classification: In classification, every object belongs to a specific *class.* This process of assigning classes is called classifying object. Classification is used where decision making and prediction is required [17]]. In healthcare study class classification is used in [9] [16] [12]. K-nearest neighbor is used for classification and regression in this input is "k" and output depends on input "k" [28]. Sampling selects data elements and creates subsets of data. Simple sampling may not perform well [17].

v. Association Rule Mining: Association rules are generated from frequent data patterns [17] [27]. In association rule, confidence and support of data is calculated [17]. On the basis of calculated support and confidence probability of occurrence of rules checked. Association rules are used in healthcare in [16] [10].

vi. Rule based Reasoning: In rule based reasoning, rules are made accordingly forward reasoning and backward reasoning are two ways of reasoning. Systems used rule based in healthcare are [25] [18].

| Table 1 Techniques and Their Use |||
|---|---|---|
| **Techniques** | **Use in healthcare** | **Count** |
| Support Vector Machine | Fall detection | 5 |
| Signal Measurement Vector | Fall detection | 1 |
| Classification and Sampling | Chronic condition, Emergency and Fall detection | 3 |
| Association rule mining | Chronic condition and Emergency detection | 2 |
| Rule Based reasoning | Chronic condition detection | 1 |
| ASP | Context awareness, Emergency and Fall Detection | 5 |
| Neural Networks | Chronic disease, Emergency and Fall detection | 1 |
| Principal Component Analysis | Fall Detection | 1 |

| Table 2 Advantages and Disadvantages |||
|---|---|---|
| **Techniques** | **Advantages** | **Disadvantages** |
| ASP | Can deal with incomplete data | Expensive to implement |
| Neural network | Can predict hidden patterns | Black box technique. Needs huge data set |
| Classification | Can classify data effectively | Define class label |
| Rule based reasoning | flexible in rule generation. More rules can be added with time | When there are more rules. More time requirement |
| Decision tree | Accuracy in output | More iterations more time |

| Table 3 Systems and accuracy measure ||||
|---|---|---|---|
| **Type** | **Systems** | **Purpose** | **Accuracy** |
| Single Purpose | CRNTC+ [9] | Emergency for seizure | **86% for seizure** |
| | Fall detection [14] | Fall detection | **96%-97%** |
| | fall detection [13] | Fall detection | **92% and 100%** |
| | Ambient Coach [15] | Movement detection | -- |
| | HUMECS [8] | Emergency detection | **91%-97%** |
| | Ontology- based [18] | Detect chronic condition | **64%** |
| | fall detection [12] | Fall detection | **91%-97%** |
| | u-care [11] | Care chart for elderly | -- |
| | m-healthcare [16]] | Emergency detection | -- |
| | Real time healthcare [10] | chronic diseases detection | **Depends on data set and classifier** |
| Multi-purpose | SINDI [26] | Emergency detection | **88%** |
| | Contexta-Care [25] | Emergency detection | -- |
| | Cyber Care[22] | Emergency detection | -- |
| | Home healthcare[23] | Emergency detection | **88.5%** |
| | Context aware monitoring [24] | Emergency detection | **60%** |

| Table 4 specifications and techniques ||||
|---|---|---|---|
| **System Type** | **System name** | **Techniques** | **Front data Collector** |
| Single Purpose | CRNTC+ 2013 | Sampling / classification | Mobile |
| | Fall detection 2014 | SMV | Wearable sensors |
| | Fall detection 2012 | MVFI, PCA, LDA | Cameras |
| | Ambient coach 2013 | Agent design DESIRE,STS | Wearable sensors |
| | HUMECS 2011 | neural network | Personal mobile host |
| | Ontology- based 2011 | Rule base reasoning, | WSN |
| | Fall detection 2011 | SVM, kNN | WSN |
| | U-Care 2013 | -- | Sensors |
| | m-healthcare 2011 | Association rules | WSN |
| | Real time healthcare 2013 | Association rules, classification | Sensors |
| Multi-Purpose | SINDI 2009 | ASP frame work | WSN |
| | Contexta-Care 2013 | based rule engine | Wearable sensors |
| | Cyber Care 2007 | ASP with predicates and logic rules | Wearable Acquisition Device |
| | Home healthcare 2010 | ASP, Decision tree | Heterogeneous |
| | Context aware monitoring 2010 | ASP | Wireless sensor network( WSN) |

vii. Answer set Programming: ASP logic is used most widely in healthcare systems. It is default reasoning method. Systems that use ASP logic are [22] [29] [23] [24] [26]. Table 1 show use of machine learning methods in healthcare systems. Frequency of technique is also shown in Table 1. From this one can observe, which technique is used most randomly. Table 2 contains advantages and disadvantages of above mention techniques. It is observed ASP performs well in multi-purpose systems and give accuracy measure "60% to 88%". While neural networks and SMV are used for single purpose systems give accuracy measure "91% to 9 7.

## IV. COMPARISON

Systems and Techniques discussed in section 3 and section 4. In this section, we compare discussed techniques and systems in terms of their accuracy, efficiency, system type and interfaces. We also analyze use of technologies and effect of data processing methods used. Table 3 shows accuracy of systems. It is to measure performance of all system. It can be observed from accuracy ratios that fall detection systems gives better accuracy rate that is between 90%-100% (100% in rare cases). Others give accuracy between 60%- 85%. Table 3 holds accuracy and specification of multi-purpose systems. Specification holds area for which system is built most of

5| Table 5 system model types and data type |||||| 
|---|---|---|---|---|---|
| Type | Methods | Model type | Set priority | Data type | Storage required |
| Single purpose | CRNTC+ | P | -- | Yes | -- |
| | Fall detection | P | -- | Yes | Yes |
| | Fall detection | -- | No | Yes | Yes |
| | Ambient coach | P | -- | Yes | Yes |
| | HUMECS | I | No | Yes | Yes |
| | Ontology based | P | -- | Yes | -- |
| | Fall Detection | P | No | Yes | -- |
| | U-care | P | No | -- | Yes |
| | m-healthcare | | Yes | Yes | Yes |
| | Real time healthcare | I | Yes | Yes | Yes |
| Multi-purpose | SINDI | P | -- | Yes | Yes |
| | Contexta-Care | UT | -- | Yes | -- |
| | Cyber Care | F | yes | Yes | Yes |
| | Home healthcare | -- | -- | Yes | Yes |
| | Context aware monitoring | -- | No | Yes | Yes |

system created are for single purpose. Table 4 holds systems data collector, techniques and year of creation of above mentioned systems. ASP is used the most for inference. It is because of its capacity to deal with incomplete data. Table 5

| Table 6 System features |||||
|---|---|---|---|---|
| Methods | filter | Human support | Updating data | Feed back |
| CRNTC+ | Yes | -- | | -- |
| Fall detection | -- | No | -- | -- |
| Fall detection | Yes | No | -- | No |
| Ambient coach | -- | No | -- | Yes |
| HUMECS | No | Yes | No | Yes |
| Ontology based context awareness | -- | -- | -- | -- |
| Automatic Fall Detection | -- | -- | -- | Yes |
| U-care | No | Yes | -- | Yes |
| m-healthcare | -- | -- | Yes | Yes |
| Real time healthcare | Yes | -- | Yes | Yes |
| SINDI | Yes | -- | Yes | Yes |
| Contexta-Care | | -- | | -- |
| Cyber Care | | -- | Yes | Yes |
| Home health care | No | No | --- | No |
| Context aware monitoring | -- | -- | -- | -- |

and Table 6 holds type of model and its storage and data requirements. We observed every system model type in categories: implemented (I), prototypical (P), Under Test (UT) and Framework (F). It is observed that most of systems created are prototypical and not applied in real world to provide healthcare services. Also most of system requires huge data storage. Healthcare systems are built with large continuous data requirement. By setting priority of data run time can be decreased. Only few system set priority of data. Table 5 and table 6 holds some features that are important for systems. Only four out of fifteen have preprocessing feature. Only two systems need human support which is good ratio. Eight out of fifteen gives feedback mean after predicting emergency it also deals with that emergency. By sending alarm, precaution or calling care-givers

## V. CONCLUSION

This research studies systems built specially for independent livings. And provide analysis of techniques used in these systems. Data is collected through sensors or cameras. Wrong settlement of sensors can affect input data. Wrong entry in online databases can also affect system output. There is need to proper pre-process data to remove noise element. ASP logic is used most widely for inference due to its feature to deal with incomplete data. Artificial Neural network (ANN) was used by one single purpose system and has shown best accuracy measure. ANN produce better results than ASP logic and other methods. It is observed that most of the system created are prototypical and not applied in real to serve health care services. Mostly systems are created for a single purpose. Mostly systems are expensive. Due to huge use of hardware some system are difficult to adopt.

## REFERENCES

[1] Marei Chan, Eric Campo, Daniel Esteve (2009) Smart homes-current features and future perspectives, "Maturitas" pp 90-97

[2] Hadi Banaee, Mobyen Uddin Ahmad and Amy Loutfi (2013), data mining for Wearable Sensors in health Monitoring systems: A review of recent trends and challenges, "sensors"

[3] Zahid Iqbal et al, Efficient Machine Learning Techniques for Stock Market Prediction, Int. Journal of Engineering Research and Applications ISSN : 2248-9622, Vol. 3, Issue 6, Dec 2013, pp.855-867

[4] Zahid Iqbal et al, "A diverse clustering Particle Swarm Optimizer for Dynamic Environment: to Locate and Track Multiple Optima" in Proc. 10th Int. Conf. IEEE Conference on Industrial Electronics and Applications, 2015, pp 1755-1760.

[5] Zahid Iqbal et al, "An Efficient indoor navigator technique to find optimal route for blinds using QR codes" in Proc. 10th Int. Conf. IEEE Conference on Industrial Electronics and Applications, 2015, pp 690-695.

[6] Zahid Iqbal et al, "A Systematic Mapping Study on OCR Techniques" Int. Journal of Computer Science and Network Security, ISSN: 2345-3397, Vol. 2, No. 1, Jan 2014.

[7] Rafia Ilyas, Zahid Iqbal, "study of hybrid approaches used for university course timetable problem"," in Proc. 10th Int. Conf. IEEE Conference on Industrial Electronics and Applications, 2015, pp 696-701.

[8] Dong-wookjang, jeongMyeongkim, "Development of a mobile health care system for rapid detection of emergent situations", Information Science and Service Science (NISS), 2011 5th International Conference on New Trends (pp.93 – 96), Volume:1

[9] CRTNC+: A smart phone-based sensor processing for prototyping personal health care application, Pervasive Computing Technologies for Healthcare (PervasiveHealth), 2013 7th International Conference , pp. 252 – 255

[10] Gouri Mohan, Sinciya P O (2013), real time healthcare system for patient with chronic disease in home and hospital environment

International Journal of Science , Engineering and technology researchVolume 2, Issue 4

[11] Eslami, M.Z, Zarghami, A. van Sinderen M. Wieringa, R,(2013),Care-giver tailoring of IT-based healthcare services for elderly at home: A field test and its results, Pervasive Computing Technologies for Healthcare (PervasiveHealth), 2013 7th International Conference Page(s):216 – 223

[12] Liang Liu, Mihail Popescu, Marjorie Skubic, Marilyn Rantz, Tarik Yardibi, Paul Cuddihy (2011), Automatic Fall Detection Based on Doppler Radar Motion Signature, 2011 5th International Conference on Pervasive Computing Technologies for Healthcare (PervasiveHealth) and Workshops

[13] David Nicholas Olivieri, Iván Gómez Conde, Xosé Antón Vila Sobrino (2012) Eigenspace-based fall detection and activity recognition from motion templates and machine learning "expert systems with applications"

[14] jin Wang, Zhongqi Zhang, Bin Li (2014), An Enhanced Fall Detection System for elderly Person Monitoring using Consumer Home Networks,

[15] Maarten F. Bobbert, Mark Hoogendoorn, Arthur J. van Soest, Vera Stebletsova, Jan Treur (2013), Ambient Support by a Personal Coach for Exercising and Rehabilitation, Human Aspects in Ambient Intelligence, Atlantis Ambient and Pervasive Intelligence Volume 8, 2013, pp 89-106,

[16] mayuri Gund, Snehal Andhalkar, prof. dipti (2011), An intelligent architecture for multi-agent based m-healthcare system, international journal of computer trends and technology

[17] JJiawei Han and Micheline Kamber(2000), data mining: concepts and techniques, Morgan Kaufmann Publishers

[18] Federica Pagnelli, Dino Giuli (2011), An ontology- based systems for context-aware and configurable services to support home based Continuous Care, IEEE transaction on information technology in biomedical, Vol. 15 No2 March 2011

[19]D. Bannach, P.lukowicz and O.Amft, (2008) "Rapid prototyping of activity recognition applications" "pervasive Computing", IEEE

[20]Centers for Disease Control and Prevention, National Center for Injury Prevention and Control. Web–based Injury Statistics Query and Reporting System (WISQARS) [online]. Accessed November 30, 2010

[21] Rosemann, M., & Recker, J. (2006). "Context-aware process design: Exploring the extrinsic drivers for process flexibility". In T. Latour & M. Petit. *18th international conference on advanced information systems engineering. proceedings of workshops and doctoral consortium*. Luxembourg: Namur University Press

[22] Mileo, A., Merico, D., & Bisiani, R. (2007). CyberCare: Reasoning about Patient's Profile in Home Healthcare. In Artificial Societies for Ambient Intelligence (pp.26-29). SSAISB Publications Officer

[23] alessandarmileo, DavideMerico (2010), A logical approach to home health care with intelligent sensor network support, the Computer journal, Vol,53 No.8

[24] Alessandarmileo, DavideMerico (2010), Support for context-aware monitoring in home healthcare, journal of Ambient Intelligence and smart Environments vol.2 49-66

[25] DavideMerico, Roberto Bisiani, Hashim Ali,(2013) Demonstrating Contextra- care: a situation- aware system for Supporting independent livings 7[th] international conference on pervasive computing technologies for health care and workshops

[26] Mileo, A. Bisiani, R.Merico, D.(2009) Pervasive wireless-sensor-networks for home healthcare need automatic reasoning, Pervasive Computing Technologies for Healthcare, 2009. PervasiveHealth 2009. 3rd International Conference pp. 1-4

[27] Tom M. Mitchell Machine Learning, McGraw-Hill Science/Engineering/Math; (March 1, 1997)

[28] Altman, N. S. (1992). "An introduction to kernel and nearest-neighbor nonparametric regression". *The American Statistician* 46 (3): 175–185.

[29] TiborBosse, Fiemke Both, Charlotte Gerritsen , Mark Hoogendoorn, Jan Treu,(2011)Methods for model-based reasoning within agent-based AmbientIntelligence applications, Knowledge-Based Systems

[30] Arvidsson, F.; Flycht-Eriksson, A. "Ontologies I" Retrieved 26 November 2008.

[31] Zuoliang Chen, Guoqing Chen, "Building an Associative Classifier based on Fuzzy Association Rules", International Journal of Computational Intelligence Systems, August, 2008

[32] M. Wooldridge, N.R. Jennings (2009) An Introduction to Multi agent systems, 2[nd] ed, John Wiley and Sons, Ltd, west Sussex, England

[33] David Isren, David Sanchez, Antonio Moreno (2010), Agents applied in healthcare, international Journal of Medical Informatics79

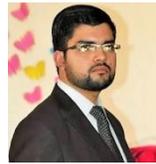
**Zahid Iqbal** received the B.S. degree in Computer Science from University of Punjab and M.S. degree in Computer Science from NUCES, FAST University, Pakistan in 2010 and 2012 respectively. He was a lecturer with Department of IT, University of Punjab, PK and since 2012, he is a Lecturer with the Faculty of Computing and IT, University of Gujrat, Punjab, PK. He is also a PhD scholar. His research interests include evolutionary algorithms, swarm intelligence, big data analytics, IoT, artificial neural networks, computational intelligence in dynamic and uncertain environments and real-world applications.

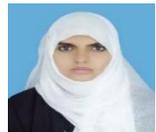
**Rafia Ilyas** received the MSc and MPhil degrees in Computer Science from University of Gujrat in 2013 and 2017 respectively. Her research interest focuses on artificial neural networks, computational intelligence, scheduling and timetabling.

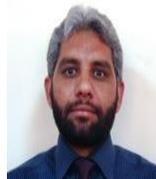
**Waseem Shahzad** received the M.S. and Ph.D. degrees in computer science from the National University of Computer and Emerging Sciences, Islamabad, Pakistan, in 2007 and 2010, respectively. Since 2010, he has been an Assistant Professor with the National University of Computer and Emerging Sciences. His current research interests include data mining, computational intelligence, machine learning, theory of computation, and soft computing. He has several publications to his credit.